\documentclass[sigconf]{acmart}
\AtBeginDocument{%
  \providecommand\BibTeX{{%
    \normalfont B\kern-0.5em{\scshape i\kern-0.25em b}\kern-0.8em\TeX}}}

\copyrightyear{2022}
\acmYear{2022}
\setcopyright{acmlicensed}
\acmConference[CIKM '22]{Proceedings of the 31st ACM International Conference on Information and Knowledge Management}{October 17--21, 2022}{Atlanta, GA, USA}
\acmBooktitle{Proceedings of the 31st ACM International Conference on Information and Knowledge Management (CIKM '22), October 17--21, 2022, Atlanta, GA, USA}
\acmPrice{15.00}
\acmDOI{10.1145/3511808.3557310}
\acmISBN{978-1-4503-9236-5/22/10}

\usepackage[utf8]{inputenc}
\usepackage{amsmath}
\usepackage{amsfonts}

\usepackage{multirow}
\usepackage{graphics}
\usepackage{balance}
\usepackage{caption}
\usepackage{subcaption}
\usepackage{algorithm}
\usepackage{algorithmic}

\newcommand{\ie}{\textit{i.e.}}
\newcommand{\eg}{\textit{e.g.}}

\title{Enhancing User Behavior Sequence Modeling by Generative Tasks for Session Search}

\author{Haonan Chen}
\author{Zhicheng Dou}
\affiliation{Gaoling School of Artificial Intelligence, Renmin University of China \state{Beijing} \country{China}}
\email{hnchen@ruc.edu.cn}
\email{dou@ruc.edu.cn}
\author{Yutao Zhu}
\affiliation{University of Montreal \city{Montreal} \state{Quebec} \country{Canada}}
\email{yutaozhu94@gmail.com}
\author{Zhao Cao}
\author{Xiaohua Cheng}
\affiliation{Poisson Lab, Huawei \state{Beijing} 
\country{China}}
\email{caozhao1@huawei.com}
\email{chengxiaohua1@huawei.com}
\author{Ji-Rong Wen}
\affiliation{Beijing Key Laboratory of Big Data Management and Analysis Methods \state{Beijing} \country{China}}
\email{jrwen@ruc.edu.cn}

\begin{document}

\begin{abstract}
Users' search tasks have become increasingly complicated, requiring multiple queries and interactions with the results. Recent studies have demonstrated that modeling the historical user behaviors in a session can help understand the current search intent. Existing context-aware ranking models primarily encode the current session sequence (from the first behavior to the current query) and compute the ranking score using the high-level representations. However, there is usually some noise in the current session sequence (useless behaviors for inferring the search intent) that may affect the quality of the encoded representations. To help the encoding of the current user behavior sequence, we propose to use a decoder and the information of future sequences and a supplemental query. Specifically, we design three generative tasks that can help the encoder to infer the actual search intent: (1) predicting future queries, (2) predicting future clicked documents, and (3) predicting a supplemental query. We jointly learn the ranking task with these generative tasks using an encoder-decoder structured approach. Extensive experiments on two public search logs demonstrate that our model outperforms all existing baselines, and the designed generative tasks can actually help the ranking task. Besides, additional experiments also show that our approach can be easily applied to various Transformer-based encoder-decoder models and improve their performance.

\end{abstract}



\maketitle

\section{Introduction} \label{sec:intro}

\begin{figure}[tbp]
\centering
\includegraphics[width=0.45\textwidth]{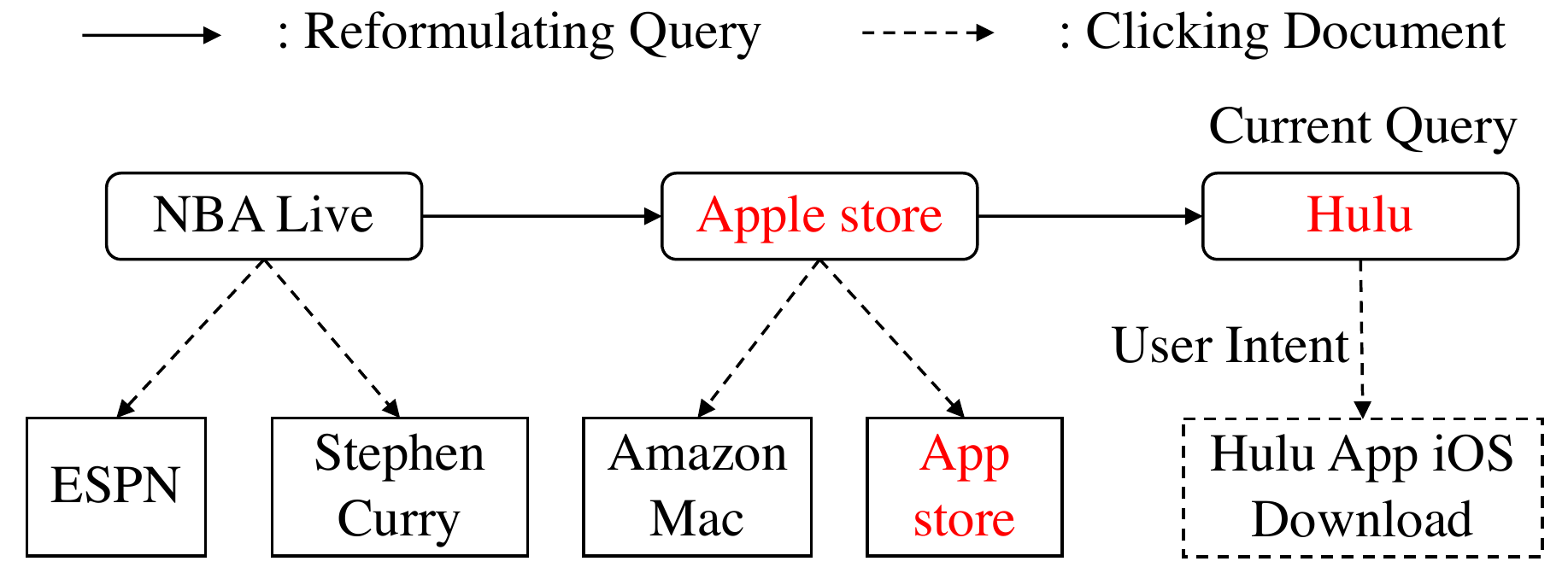}
\caption{An example of session context that contains noise. The queries and documents that we believe can help infer the user's search intent are marked with the color \textcolor{red}{red}.}
\label{fig:noise_example} 
\end{figure}

With the development of search engines, users' information needs have become increasingly complex. 
A user usually issues several queries and examines some documents to complete a search task.
These user behaviors (\eg, issued queries and clicked documents) that occur during a relatively brief period are referred to as a \emph{search session}~\cite{coca, cars, jones2008, Wang2013}. 
Modeling the session context has been demonstrated to be beneficial for understanding search intent~\cite{coca, cars}. 

Several early studies have attempted to model session context based on statistical techniques, which inevitably neglect some valuable features~\cite{Shen2005, VanGysel2016, Bennett2012, White2013}. 
With the emergence of deep learning, many neural context-aware ranking models have been proposed~\cite{cars, ricr, hba, DBLP:conf/cikm/ZhouD0W21, mmtensor}. 
They use recurrent neural networks (RNNs) to encode user behaviors into latent representation~\cite{cars, mmtensor,ricr}, or pre-trained language models (PLMs), such as BERT~\cite{bert}, to get a context-aware representation of the session sequence~\cite{hba, coca}.
This representation is used to compute ranking scores.
However, the current session sequence (from the beginning to the current query) may contain some useless information (\ie, noise) that could cause the encoders (\eg, BERT or RNNs) to misinterpret the real user intent.
Figure~\ref{fig:noise_example} illustrates an example where a user has issued several queries and examined some documents. 
The current query is ``Hulu'', and the user is trying to download the iOS version of the Hulu application. 
Evidently, the previous query ``NBA live'' is useless for inferring the current search intent,
and simply encoding such noise may degrade the sequence's representation.
Unfortunately, the quality of the session representation, \ie, whether the real user intent has been encoded, has received little attention in existing studies.

A straightforward way to tackle this issue is to apply the auto-encoder technique.
Auto-encoder has an encoder-decoder structure, where the decoder is used to recover the input sequence based on the representation computed by the encoder. The encoder is thereby trained to capture the most important information from the input sequence.
However, applying auto-encoder to session search is non-trivial in \textbf{three aspects}: 
{(1)} Generating the whole session sequence is challenging.
Typically, a session consists of many queries and documents, which are too long for the decoder to recover. 
Besides, most auto-regressive decoders, including those pre-trained on large-scale corpora (\eg, GPT~\cite{gpt}), are incapable of modeling the relevance between queries and documents~\cite{prop}.
{(2)} There is noise in the session sequence. 
As indicated above, a session sequence usually contains user behaviors irrelevant to the current information need. Thus, the decoder should not generate all behaviors in the session sequence without differentiation. 
{(3)} The user behaviors that represent the real search intent may be implicit in the current session sequence.
A user's current information need is often complicated and cannot be described clearly by the user (or understood by the search engine).
For example, the behavior reflecting the user's information need may be a future query in the session or a similar query issued by a different user in another session (\ie, another user can successfully address the same information need while this user cannot).
Under this circumstance, simply recovering the current session sequence is ineffective.

To address these problems, we employ an encoder-decoder structure and design several generative tasks specifically for session search to assist the encoder in inferring the search intent more accurately. 
Specifically, we design \textbf{three generative tasks}:

\textbf{\underline{Task 1}}: Predicting future queries. 
As the session progresses, the user becomes more explicit about their actual information need.
Thus, subsequent queries within the same session can more accurately reflect the search intent.

\textbf{\underline{Task 2}}: Predicting future clicked documents. 
In addition to future queries, we also consider future user clicks because the documents usually contain more detailed information than keyword-based queries.

\textbf{\underline{Task 3}}: Predicting a supplemental query.
As explained in the third problem above, some queries in other users' sessions may be helpful in understanding the current search intent.

All of these generative targets are more accurate (or supplemental) descriptions of the current search intent.
Therefore, only if the encoder has successfully encoded the user's search intent into the representation can the decoder predict these sequences using the representation of the current user behavior sequence.
Besides, our designed generative tasks can \textbf{address the three aforementioned challenges as follows}:
For the first problem, Task 1\&2 attempt to generate the future queries and documents separately, so avoiding generating long sequences or modeling relevance between queries and documents, making the generation easier.
For the second problem, we explore many potential generation targets and propose these three tasks that can actually help the encoder infer actual search intent.
Our experiments in Section~\ref{subsec:explore_generation} will show the effectiveness of these generative tasks.
For the third problem, all these tasks try to predict future sequences (or a supplemental query), \ie, information that is not in the current sequence.

We propose to jointly learn the ranking and generative tasks by an encoder-decoder structured approach.
Specifically, we attempt to use future sequences and a supplemental query as generation targets to enhance the encoder's ability to represent session context.
We call our model \textbf{ASE} -- \textbf{A}uto-\textbf{S}ession-\textbf{E}ncoder, which is based on a pre-trained BART~\cite{bart}.
Experimental results on two public search logs (AOL~\cite{aol} and Tiangong-ST~\cite{tiangong}) show that ASE outperforms existing methods, which demonstrates its effectiveness. 
Moreover, the consistent performance improvements on top of different Transformer-based encoder-decoder models demonstrate our approach's effectiveness and wide applicability.

To summarize, the contributions of this work are as follows:

(1) We propose Auto-Session-Encoder, which employs several generative tasks to explicitly enhance the ability to encode a user behavior sequence under an encoder-decoder framework. 

(2) We design three generative tasks to utilize the future sequences and a supplemental query to train a better representation of the current session sequence.
Experimental results demonstrate the effectiveness of the generative tasks. 

(3) We demonstrate that our model can be easily adapted to various Transformer-based encoder-decoder models other than BART, indicating its wide applicability.

\section{Related Work} \label{sec:RW}

\subsection{Session Search}

There are already some traditional approaches that utilize session context to infer search intent~\cite{Shen2005, VanGysel2016, Bennett2012, White2013, carterette2016evaluating}.
Specifically, Shen et al.~\cite{Shen2005} used statistical language models to combine session context and the current query for better ranking performance.
Van Gysel et al.~\cite{VanGysel2016} explored lexical query modeling for session search. 
They found that specialized session search methods are more suitable for modeling long sessions than naive term weighting methods.

With the emergence of deep learning, researchers have focused on designing neural context-aware ranking models~\cite{cars, mmtensor, hba, ricr, hqcn, coca}.
Specifically, Ahmad et al.~\cite{mmtensor, cars} encoded queries and documents using RNNs and attention mechanism.
Then they jointly learned the ranking task and query suggestion task.
Qu et al.~\cite{hba} concatenated the current session sequence and put them into a BERT encoder.
Then they applied a hierarchical behavior-aware attention module over the BERT encoder to get high-level representations for ranking.
Zuo et al.~\cite{hqcn} modeled multi-granularity historical query change. 
They obtained multi-level representations of the session using Transformer-based encoders.
Chen et al.~\cite{ricr} integrated representation and interaction.
They encoded the session history into a latent representation and used it to enhance the current query and the candidate document. 
Then they captured the interaction-based information between the enhanced query and the candidate document.
Zhu et al.~\cite{coca} utilized data augmentation and contrastive learning to pre-train a BERT encoder that can represent the session sequence better.
Most existing models use an encoder to model the current session sequence and obtain high-level representations of the sequence to compute ranking scores.
However, because of noise in the session, the representations may fail to encode the user's actual search intent.
Our model aims to enhance user behavior sequence modeling by multiple designed generative tasks that attempt to utilize future sequences and a supplemental query.
By this, we attempt to explicitly ensure the actual search intent has been encoded into the high-level representations.

\subsection{Generative Tasks for IR}

There are already some works trying to utilize generative tasks to improve retrieval performance~\cite{gdmtl, GAR4QA, cars, mmtensor, losnet}.
Liu et al.~\cite{gdmtl} demonstrated that generative tasks can make retrieval modeling more generalized.
Mao et al.~\cite{GAR4QA} showed that generating heuristically discovered relevant contexts for queries can improve their retrieval and QA results.
Cheng et al.~\cite{losnet} utilized the next query prediction task to help personalized re-ranking.
Ahmad et al.~\cite{mmtensor, cars} illustrated that the query suggestion task could improve the ranking quality of session search.
Though they have already demonstrated the effectiveness of predicting the next query, there are more generation targets to be explored.
Specifically, we find that predicting future queries (not only the next query), predicting future clicked documents, and predicting a supplemental query can all help model user behavior sequences.
After exploring various potential generative targets, we design multiple generative tasks specifically for session search (Section~\ref{subsec:generate}) to help model the current session context.

\section{Auto-Session-Encoder} \label{sec:method}

Session search aims to utilize the user behavior sequence to rank candidate documents.
Most existing models use an encoder to model session context and get high-level representations of the sequence to compute ranking scores.
However, the representations may lack the information of the user's actual search intent because of noisy user behaviors.
Our model aims to enhance the ability of the encoder with three designed generative tasks that attempt to utilize the information of future sequences and a supplemental query.
By this, we try to help the encoder to encode the actual search intent into the high-dimensional representations of the session sequence.

\subsection{Problem Definition}

Before shedding light on our proposed model, we will state some notations about session search.
Suppose a query $q_i$ has $M$ clicked documents $D_i = \{d_{i,1}, d_{i,2},\cdots, d_{i,M}\}$.
Following~\cite{coca, hba, hqcn}, we keep the first clicked document for each historical query to construct the sequence.
Then the current session sequence $S$ when the user is issuing the $n$-th query $q_n$ can be denoted as:
\begin{equation}
    S = \{(q_1, d_1), (q_2, d_2), \cdots , (q_n)\}. \notag
\end{equation}

The goal of session search (or context-aware document ranking) is to model the contextual information to obtain the ranking scores of the candidate documents $D_c$ and rank them accordingly. 
We will focus on how to get the score of a candidate document ($d_c$) in the rest of the paper.
Note that the current session sequence $S$ only contains the historical and present user behaviors when a user is issuing $q_n$. 
However, we will utilize future sequences and a supplemental query as generation targets while training.

\subsection{Overall Structure} \label{subsec:overview}

\begin{figure*}[tbp]
\centering
\includegraphics[width=0.8\textwidth]{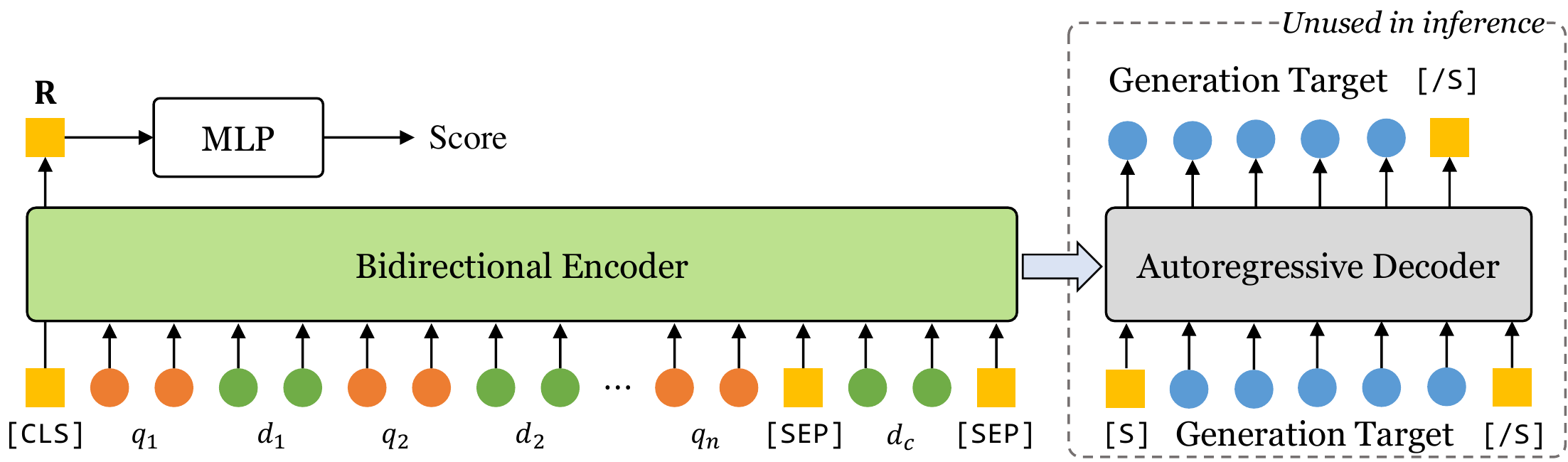}
\caption{The diagram of ASE. 
\textbf{Training}: For ranking, ASE concatenates the current user behavior sequence and puts it into the encoder. 
Then ASE makes the output of the ``\texttt{[CLS]}'' token go through an MLP to get the ranking score of $d_c$. For each generative task, ASE takes the generative target as the generation label of the decoder. 
With the ranking loss and generation losses ready, ASE jointly learns these tasks to enhance the encoder. 
\textbf{Inference}: ASE only uses the encoder to rank the candidate documents.}
\label{fig:model} 
\end{figure*}

In this part, we will introduce the overall structure of ASE. ASE jointly learns the ranking task and the generative tasks as follows:

(1) \textbf{Ranking.} 
As shown in the left part of Figure~\ref{fig:model}, we attempt to compute the ranking score of the candidate document $d_c$ for the ranking task.
To model the session context, ASE first concatenates the session sequence $S$ with $d_c$ and puts it into the encoder.
Then ASE gets the output of the ``\texttt{[CLS]}'' token as the high-dimensional representation.
Finally, we apply a linear projection on this representation to get the ranking score of $d_c$ (Section~\ref{subsec:rank}).

\begin{figure}[tbp]
\centering
\includegraphics[width=0.48\textwidth]{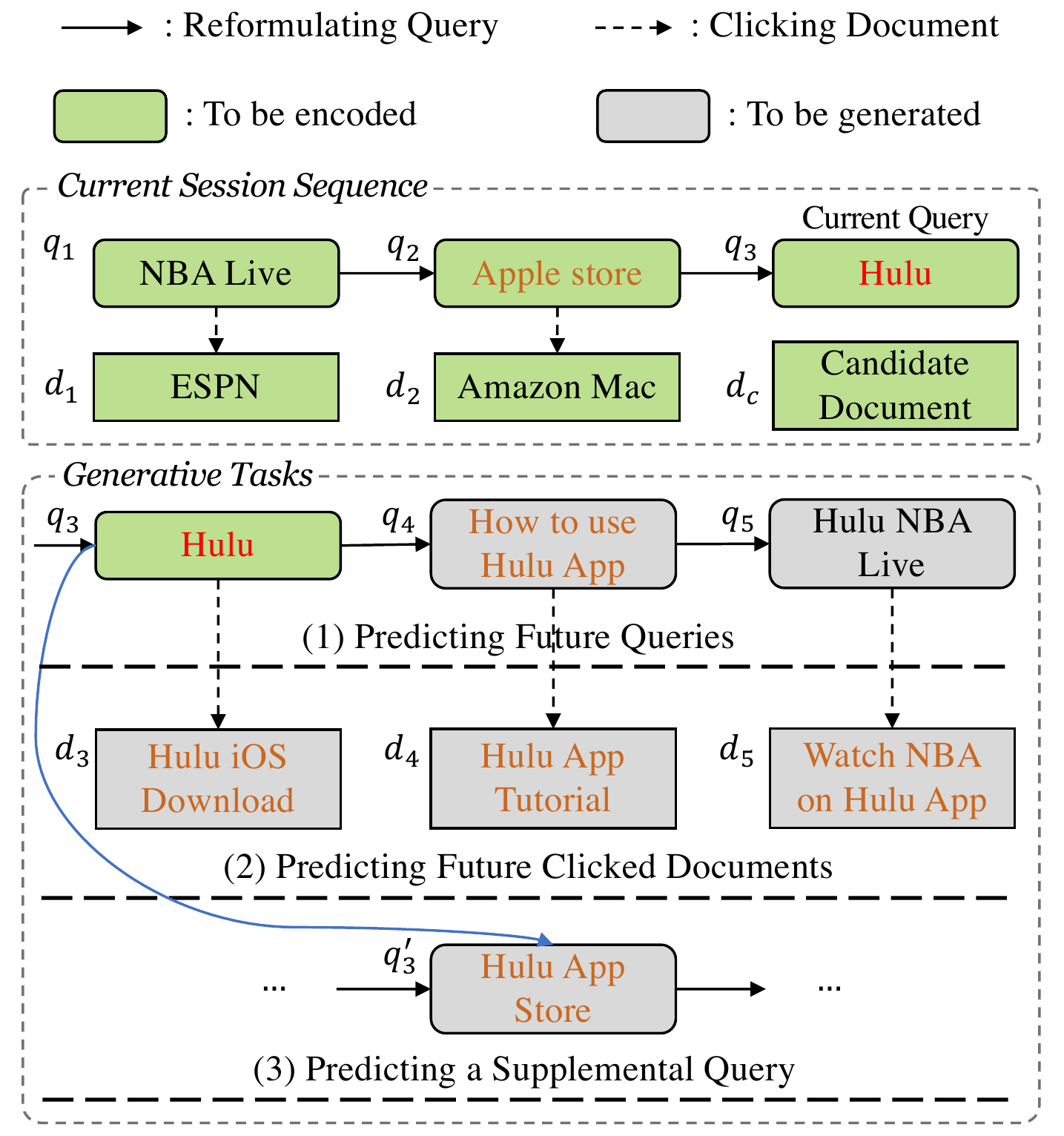}
\caption{The three generative tasks designed for session search. 
They are only used when training.
We take a session that has five query-document pairs as an example.
Suppose $q_3$ is the current query, then our goal is to utilize the information of future sequences and a supplemental query to model the current user behavior sequence.
The queries and documents that we believe can help infer the user's current search intent are marked with the color \textcolor{red}{red}.
}
\label{fig:gen_tasks}
\end{figure}

(2) \textbf{Generation.}
As shown in Figure~\ref{fig:gen_tasks} and the right part of Figure~\ref{fig:model}, we aim to enhance the ability of the encoder using the decoder and three generative tasks (Section~\ref{subsec:generate}).
These generative tasks are comprised of
(i) predicting future queries,
(ii) predicting future clicked documents, and
(iii) predicting a supplemental query. 

Finally, by jointly learning the ranking task and the generative tasks (Section~\ref{subsec:loss}), the encoder can model user behavior sequences better and learn representations that contain actual search intent. Note that these generative task are only used in the training stage, for enhancing the representation ability of the encoder. \textbf{At inference time, we will only use the enhanced encoder to score the candidate documents}.

In this work, we choose the pre-trained language model BART~\cite{bart} as ASE's backbone because:
(1) BART is a Transformer-based~\cite{transformer} encoder-decoder model with a bidirectional (BERT-like) encoder and an autoregressive (GPT-like) decoder. 
We can naturally implement our ranking and designed generative tasks on this model.
(2) BART utilizes self-supervised pre-training, which makes it perform very well on many generative tasks and do not reduce performance on discriminative tasks~\cite{bart}. 
(3) BART-base model uses six layers in the encoder and decoder, respectively, which makes it contain the comparable number of parameters as BERT-base model (twelve layers in the encoder).
Besides, the number of training steps and the data used for pre-training BART is the same as BERT. 
Thus, we choose BART as ASE's backbone for fair comparisons with BERT-based baseline models~\cite{hba,coca}.
In addition, as demonstrated in Section~\ref{subsec:other_plms}, we can easily apply ASE to other Transformer-based encoder-decoder models.

\subsection{Modeling the Current Session Sequence} \label{subsec:rank}

In this section, we will illustrate how ASE uses the encoder to model session context. 
As shown in the left part of Figure~\ref{fig:model}, ASE first puts the concatenated sequence into the encoder and gets a high-level representation from the ``\texttt{[CLS]}'' token. 
Then ASE makes it go through a linear projection to get the ranking score.

Following previous works~\cite{hba, coca}, we treat the ranking task as a sequence pair classification problem. 
Given the session sequence $S$, we consider the current user behaviors as one sequence $\{q_1, d_1, q_2, d_2, ... , q_n\},$ and the candidate document to be scored as another sequence $\{d_c\}$.
After adding some special tokens, the constructed input sequence is as follows:
\begin{equation} 
    I = \texttt{[CLS]}q_1\texttt{[EOS]}d_1\texttt{[EOS]} \cdots q_n\texttt{[EOS]}\texttt{[SEP]}d_c\texttt{[EOS]}\texttt{[SEP]}, \notag
\end{equation}
where ``\texttt{[CLS]}'' is the classification token, ``\texttt{[EOS]}'' is used to identify the end of each query and document, ``\texttt{[SEP]}''
is used to separate the sequence pair for classification.

Then ASE makes $I$ go through the encoder and takes the output of ``\texttt{[CLS]}'' token as the high-level representation:
\begin{equation}
    \textbf{R} = {\rm Encoder} (I)_{\texttt{[CLS]}}.
\end{equation}
Finally, we use a multi-layer perceptron (MLP) on this representation to get the ranking score:
\begin{equation}
    {\rm Score}(d_c) = {\rm MLP} (\textbf{R}).
\end{equation}

\subsection{Enhancing Encoder with Generative Tasks} \label{subsec:generate}

As stated in Section~\ref{sec:intro}, the current session sequence $S$ may contain some noise that is irrelevant to the current search intent, which may affect the quality of the session representation $\textbf{R}$.
Thus, we propose to enhance the encoder's ability with a decoder and three generative tasks that are designed for session search. 
Specifically, we attempt to utilize the information of future sequences and a supplemental query.
This can help the encoder to encode the actual search intent into $\textbf{R}$.

As presented in Figure~\ref{fig:gen_tasks}, we take a session that has five query-document pairs as an example to illustrate the three designed generative tasks.
Suppose during training, $q_3$ ``Hulu'' is the current query, and there are some candidate documents to be ranked. 
Then the current user behavior sequence is $\{q_1, d_1, q_2, d_2, q_3\}$.
Let us suppose that the user is trying to download the application Hulu from App Store.
However, the current sequence has some noise (queries and documents that are not marked \textcolor{red}{red}). 
For example, $q_1$ ``NBA Live'' may mislead the search system to infer that the user is trying to watch live NBA games on the Hulu website.

Since the current session is noisy, the encoder may have trouble encoding the actual search intent into $\textbf{R}$.
We propose to utilize generative tasks to help the encoding of the current sequence $S$ during training.
Specifically, we design three generative tasks for session search as follows:

(1) \textbf{Predicting Future Queries.}
The user often tries to issue a new query when the current one does not satisfy her information need.
Besides, as the searching progresses, the user has a more explicit understanding of the search task, and the quality of the queries they issue is also getting higher.
Thus, the subsequent queries in the session can often represent the search intent better than the current query.
For example, if ASE can predict $q_4$ ``How to use Hulu App'' given the information of $S$, then we believe the encoder has encoded the user's search intent (download the Application Hulu from App Store) into the high-level representations.
To utilize the information of the future queries, we concatenate them as a generation target of the decoder as follows:
\begin{equation}
    GT_1 = q_{n+1}{\texttt{[SEP]}}q_{n+2}{\texttt{[SEP]}}\cdots q_{n+w}{\texttt{[SEP]}}, \notag
\end{equation}
where $w$ is the prediction window size that controls how many subsequent behaviors of $q_n$ we want to generate.

(2) \textbf{Predicting Future Clicked Documents.} 
Similar to the reasons above, the future clicked documents (especially the current query's click, which we also consider a future clicked document) often contain valuable information about the user's search interest.
Besides, we can usually get more detailed information from documents
than queries since many queries contain only keywords.
For example, $d_3$ ``Hulu iOS Download'' is obviously a more specific and accurate version of $q_3$. It would be great if the decoder could predict this information from the high-dimensional representations of $S$.
Thus, we believe the future clicked documents can also help the encoder infer the search intent, and we utilize their information as follows:
\begin{equation}
    GT_2 = d_{n}{\texttt{[SEP]}}d_{n+1}{\texttt{[SEP]}}\cdots d_{n+w-1}{\texttt{[SEP]}}. \notag
\end{equation}
We will notice that the generation starts at the current clicked document $d_n$ since it is considered a future sequence.

(3) \textbf{Predicting a Supplemental Query.}
Many queries contain only keywords, making them hard to be understood, especially when the session sequence is noisy.
Besides, if the user fails to address her search task, then the user behavior that can represent her search intent may be implicit in the current session.
For example, the query from another session $q_3'$ ``Hulu App Store'' can supplement our model's understanding of $S$.
Following previous works~\cite{ricr, psgan}, we attempt to find a query to supplement the information of the current query, which can make our model more robust.
We treat the training data as the query database, and for each query, we mine one query that we believe contains supplemental information from the database. 
We use the equation suggested by Chen et al.~\cite{ricr} to measure the supplemental rate of the query we choose:
\begin{equation}
    {\rm sup}(q_n',q_n) = {\rm spe}(q_n',q_n) + {\rm sim}(q_n',q_n),
\end{equation}
where $q_n'$ is the candidate query in the database, ${\rm sup}(q_n',q_n)$ is its supplemental rate; ${\rm spe}(q_n',q_n) = \frac{{\rm len}(q_n') - {\rm len}(q_n)}{{\rm len}(q_n)}$ when every word of $q_n$ appear in $q_n'$, otherwise it is 0. 
This component computes the specificity between $q_n'$ and $q_n$; ${\rm sim}(q_n',q_n) $ is the similarity between $q_n'$ and $q_n$, which is computed by the python class SequenceMatcher.\footnote{\url{https://docs.python.org/3/library/difflib.html}}

We choose the query $q_n'$ that has the highest supplemental rate to be our last generation target:
\begin{equation}
    GT_3 = q_n'{\rm \texttt{[SEP]}}. \notag
\end{equation}
Different from $GT_1$ and $GT_2$, we only use one sequence here as the generation target. 
This is because the queries mined from other sessions often represent different information needs, and we do not want to confuse our model with too many different topics.

For those queries and documents in $GT_1$ and $GT_2$ that are empty (lacking future information or recording error in the datasets), we use ``[empty\_q]'' and ``[empty\_d]'' to pad them respectively.

With these generation targets ready, we treat them as different generative tasks and train the decoder to generate them separately during training.
If the decoder could predict these targets from the high-dimensional representations, we believe the encoder has successfully encoded the actual search intent.
Extensive experiments in Section~\ref{subsec:explore_generation} show the effectiveness of the generative tasks.

\subsection{Optimizing Ranking and Generation Jointly}\label{subsec:loss}

In this part, we will learn the above tasks jointly by a multi-task technique. 
Following~\cite{gdmtl}, in order to automatically balance the importance of these tasks, we apply a variation~\cite{uncertainty2} of the Uncertainty~\cite{uncertainty} technique to learn the weights:
\begin{equation}
    \mathcal{L} = \frac{\mathcal{L}_R}{2\tau_r^2} + {\rm log} (\tau_r^2+1) + \sum_{g \in G} \left( \frac{\mathcal{L}_g}{2\tau_g^2} + {\rm log} (\tau_g^2+1) \right),
\end{equation}
where $\mathcal{L}_R$ is the ranking loss, $\mathcal{L}_g$ is one of the generation losses, $\tau$s are tunable parameters that represent the uncertainty.

The intuition here is that if the value of a loss is too high, then its corresponding uncertainty will also increase to reduce its contribution to the main loss.
More details of this technique can be found in its original paper~\cite{uncertainty}.

To implement the ranking loss of $q_n$ ($\mathcal{L}_R(q_n)$), we apply a pair-wise ranking function hinge loss as follows:
\begin{equation}
    \mathcal{L}_R(q_n) = \sum_{(d^+_c,d^-_c) \in D_c}^{} 
    {\rm max} \left( 0, \alpha - Score(d^+_c) + Score(d^-_c) \right), 
\end{equation}
where $\alpha$ is a hyperparameter of margin, which is set as 1 for binary classification task, $D_c$ is the candidate documents of $q_n$, $d^+_c$ is a clicked document, and $d^-_c$ is a skipped document.
We attempt to use this loss to train ASE to re-rank relevant documents higher than irrelevant ones.

For each generation target $GT$, its generation loss ($\mathcal{L}_g(GT)$) is implemented as the negative log-likelihood of predicting $GT$ based on $S$ and $d_c$:
\begin{equation}
    \mathcal{L}_g(GT) = -\sum^{|GT|}_{j=1} 
    {\rm log} (Pr(w_j|w_{1:j-1}, S, d_c)).
\end{equation}

\begin{table}[t!]
    \centering
    \small
    \caption{Statistics of AOL and Tiangong-ST.}
    \begin{tabular}{lrrrrrr}
    \toprule
        \textbf{AOL} & \textbf{Training} & \textbf{Validation} & \textbf{Test} \\
    \midrule
        \# Session & 219,748 & 34,090 & 29,369 \\
        \# Query  & 566,967 & 88,021 & 76,159 \\
        Average Session Length & 2.58 & 2.58 & 2.59 \\
        \# Candidate per Query & 5 & 5 & 50 \\
        Average Query Length & 2.86 & 2.85 & 2.9 \\
        Average Document Length & 7.27 & 7.29 & 7.08 \\ 
        Average \# Click per Query & 1.08 & 1.08 & 1.11 \\
    \midrule
        \textbf{Tiangong-ST} & \textbf{Training} & \textbf{Validation}  & {\textbf{Test}} \\
    \midrule
        \# Session & 143,155 & 2,000 & {2,000}\\
        \# Query  & 344,806 & 5,026 &  {6,420}\\
        Average Session Length & 2.41 & 2.51 & {3.21}\\
       \# Candidate per Query & 10 & 10 & {10}\\
        Average Query Length & 2.89 & 1.83 & {3.46}\\
        Average Document Length & 8.25 & 6.99 & {9.18}\\
        Average \# Click per Query & 0.94 & 0.53 & {3.65}\\
    \bottomrule
    \end{tabular}
    \label{tab:dataset}
\end{table}

\section{Experimental Setup} \label{sec:setting}

\subsection{Datasets and Evaluation Metrics}

\subsubsection{Datasets}

We conduct our experiments on AOL search log~\cite{aol} and Tiangong-ST search log~\cite{tiangong}. 
They are both public large-scale search logs.
We have also considered MS MARCO Conversational Search dataset.\footnote{\url{https://github.com/microsoft/MSMARCO-Conversational-Search}}
However, the sessions of this dataset are artificial, and we want to study actual user behaviors from real-world search logs.
Therefore, we do not use this dataset and stick to AOL and Tinagong-ST which are widely used in existing works.

We process the AOL search log following Ahmad et al.~\cite{cars}. 
Each query of the training and validation sets contains five candidate documents, and each one of the test set contains 50 candidates that are retrieved by the BM25 algorithm~\cite{bm25}.

Tiangong-ST~\cite{tiangong} is collected from a Chinese commercial search engine.
For the last query of each in the test set, its candidate documents have human-annotated relevance labels (0 to 4).
As suggested by the original paper of this dataset~\cite{tiangong}, we will use the queries that have relevance labels when testing.
For more details on this dataset, please refer to~\cite{tiangong}.

Following previous works~\cite{cars, hba, ricr, hqcn, coca}, we only use the title of each document as its content. The statistics of these two datasets are presented in Table~\ref{tab:dataset}.

\subsubsection{Evaluation Metrics}

Following previous works~\cite{cars, hba, ricr, hqcn, coca}, we use Mean Average Precision (MAP), Mean Reciprocal Rank (MRR), and Normalized Discounted Cumulative Gain (NDCG) at position $k$ (NDCG@$k$, $k=1,3,5,10$) as metrics.
We use TREC's evaluation tool (trec\_eval)~\cite{pytrec} to compute all evaluation results.

\subsection{Baselines} \label{subsec:baselines}

Following previous works~\cite{ricr, hqcn, coca}, we compare ASE with two kinds of baselines: 

(1) \textbf{Ad-hoc ranking models} only use the information of $q_n$ and $d_c$ to get the ranking score.

$\bullet$ \textbf{BM25}~\cite{bm25} is a traditional ranking algorithm based on the probabilistic retrieval framework. It treats the relevance between $q_n$ and $d_c$ as a probability problem.
$\bullet$ \textbf{ARC-I}~\cite{arci} obtains the representations of $q_n$ and $d_c$ by  convolutional neural networks (CNNs) and treats the semantic similarity as $d_c$'s relevance to $q_n$.
$\bullet$ \textbf{ARC-II}~\cite{arci} obtains the word-level interaction-based information of $q_n$ and $d_c$ using 2D-CNNs.
$\bullet$ \textbf{KNRM}~\cite{knrm} utilizes soft matching signals by kernel pooling on the interaction matrix of $q_n$ and $d_c$.
$\bullet$ \textbf{Duet}~\cite{duet} integrates both interaction-based and representation-based features to score $d_c$.

(2) \textbf{Context-aware ranking models} attempt to understand the search intent by modeling session context.

$\bullet$ \textbf{CARS}~\cite{cars} uses RNNs and the attention mechanism to encode user behaviors and sequential information of session history into latent representations.
It computes the ranking score and jointly suggests useful queries to the user based on these representations.
$\bullet$ \textbf{HBA-Transformers}~\cite{hba} concatenates $S$ with $d_c$ and puts them into a BERT encoder. Then it applies a hierarchical behavior-aware attention module over the BERT encoder to model interaction-based information at different levels.
$\bullet$ \textbf{HQCN}~\cite{hqcn} attempts to model multi-granularity historical query change. It also introduces the query change classification task to help rank candidates.
$\bullet$ \textbf{RICR}~\cite{ricr} integrates representation and interaction. Instead of making every two behaviors interact with each other, it first uses the representation of session history to enhance $q_n$ and $d_c$. Then it makes the enhanced $q_n$ and $d_c$ interact on the word level. 
$\bullet$ \textbf{BERT}~\cite{bert} and \textbf{BART}~\cite{bart} are the vanilla versions of BERT-base and BART-base. We include these two as baselines to demonstrate that it is fair (Section~\ref{subsec:overview}) to compare our BART-based model ASE to BERT-based baselines (HBA, COCA). When fine-tuning these two models, we simply concatenate $S$ with $d_c$ and put it into the encoder (we do not use the decoder of BART). Then we make the output of \texttt{[CLS]} go through an MLP to get the ranking score.
$\bullet$ \textbf{COCA}~\cite{coca} utilizes data augmentation and contrastive learning to pre-train a BERT encoder that can represent the session sequence better. It is the state-of-the-art model which has been demonstrated effective in NTCIR-16 Session Search Track~\cite{rucir_ntcir, chen2022overview}.

\subsection{Implementation Details} \label{subsec:implementation}

For AOL, we use the BART-base model provided by the authors of~\cite{bart} on Huggingface.\footnote{\url{https://huggingface.co/facebook/bart-base}}
For Tiangong-ST, we use the Chinese BART-base model provided by the authors of~\cite{chineseBart} on Huggingface.\footnote{\url{https://huggingface.co/fnlp/bart-base-chinese}}
Following T5~\cite{t5}, we use a unique task identifier at the beginning of the input sequence for each task.
Following previous works~\cite{hba, coca}, we truncate the sequence from the head if its length exceeds the maximum length.

For details of the instantiations, one can refer to our code.\footnote{\url{https://github.com/haon-chen/ASE-Official}}

\section{Results and Analysis} \label{sec:result}

\subsection{Overall Results} \label{subsec:overall result}

\begin{table}[t!]
    \centering
    \small
    \caption{Overall results on AOL and Tiangong-ST.
    ``$\dag$'' and ``$\ddag$'' denote the result is significantly worse than our  ASE in t-test with $p$-value $<$ 0.01 and $p$-value $<$ 0.05 respectively.
    The best performance is in bold.} 
    \begin{tabular}{lcccccl}
    \toprule
        \multicolumn{7}{c}{\textbf{AOL}} \\
         Model & NDCG@1 & @3 & @5 & @10 & MAP & MRR \\ 
        
        \cmidrule(lr){1-7}
        
        \multicolumn{7}{l}{Ad-hoc Ranking Models}  \\
        
        \cmidrule(lr){1-7}
        
        {BM25} &  {0.1195$^\dag$} &  {0.1862$^\dag$} &  {0.2136$^\dag$}
        &  {0.2481$^\dag$} &  {0.2200$^\dag$} &  {0.2271$^\dag$} \\
        
        ARC-I & 0.1988$^\dag$ & 0.3108$^\dag$ & 0.3489$^\dag$ & 0.3953$^\dag$ & 0.3361$^\dag$ & 0.3475$^\dag$ \\
        
        ARC-II & 0.2428$^\dag$ & 0.3564$^\dag$ & 0.4026$^\dag$ & 0.4486$^\dag$ & 0.3834$^\dag$ & 0.3951$^\dag$ \\
        
        KNRM & 0.2397$^\dag$ & 0.3868$^\dag$ & 0.4322$^\dag$ & 0.4761$^\dag$ & 0.4038$^\dag$ & 0.4133$^\dag$ \\
        
        Duet & 0.2492$^\dag$ & 0.3822$^\dag$ & 0.4246$^\dag$ & 0.4675$^\dag$ & 0.4008$^\dag$ & 0.4111$^\dag$  \\
        
        \cmidrule(lr){1-7}
        
        \multicolumn{7}{l}{Context-aware Ranking Models}  \\
        
        \cmidrule(lr){1-7}
        
        CARS & 0.2816$^\dag$ & 0.4117$^\dag$ & 0.4542$^\dag$ & 0.4971$^\dag$ & 0.4297$^\dag$ & 0.4408$^\dag$ \\
        
        HBA & 0.3773$^\dag$ & 0.5241$^\dag$ & 0.5624$^\dag$ & 0.5951$^\dag$ & 0.5281$^\dag$ & 0.5384$^\dag$ \\ 
        
        RICR & 0.3894$^\dag$ & 0.5267$^\dag$ & 0.5648$^\dag$ & 0.5971$^\dag$ & 0.5338$^\dag$ & 0.5450$^\dag$ \\ 
        
        HQCN & 0.3990$^\dag$ & 0.5441$^\dag$ & 0.5783$^\dag$ & 0.6070$^\dag$ & 0.5448$^\dag$ & 0.5549$^\dag$ \\
        
        BART & {0.3908$^\dag$} & {0.5414$^\dag$} & {0.5797$^\dag$} & {0.6108$^\dag$} & {0.5450$^\dag$} & {0.5551$^\dag$} \\
        
        BERT & {0.3990$^\dag$} & {0.5440$^\dag$} & {0.5818$^\dag$} & {0.6123$^\dag$} & {0.5471$^\dag$} & {0.5572$^\dag$} \\
        
        COCA & {0.4024}$^\dag$ & {0.5478}$^\dag$ & {0.5849}$^\dag$ & {0.6160}$^\dag$ & {0.5500}$^\dag$ & {0.5601}$^\dag$ \\

 {ASE} & \textbf{0.4144} & \textbf{0.5682} & \textbf{0.6007} & \textbf{0.6283} & \textbf{0.5650} & \textbf{0.5752} \\
 
        \midrule
        \hline
        \multicolumn{7}{c}{\textbf{Tiangong-ST}} \\
        Model & NDCG@1 & @3 & @5 & @10 & MAP & MRR \\
        
        \cmidrule(lr){1-7}
        
        \multicolumn{7}{l}{Ad-hoc Ranking Models}  \\
        
        \cmidrule(lr){1-7}
        
        {BM25} &  {0.6029$^\dag$} &  {0.6646$^\dag$} &  {0.7072$^\dag$} &  {0.8541$^\dag$} & {0.7837$^\dag$} & {0.8225$^\dag$} \\
        
        ARC-I & 0.7088$^\dag$ & 0.7087$^\dag$ & 0.7317$^\dag$ & 0.8691$^\dag$ & 0.8580$^\ddag$ & 0.9159$^\dag$ \\
        
        ARC-II & 0.7131$^\dag$ & 0.7237$^\dag$ & 0.7379$^\dag$ & 0.8732$^\dag$ & 0.8611$^\ddag$ & 0.9227$^\dag$ \\
        
        KNRM & 0.7198$^\dag$ & 0.7421$^\dag$ & 0.7660$^\dag$ & 0.8857$^\ddag$ & {0.8683} & 0.9130$^\dag$ \\
        
        Duet & 0.7577$^\ddag$ & 0.7354$^\dag$ & 0.7548$^\dag$ & 0.8829$^\ddag$ & 0.8663 & 0.9273$^\ddag$ \\
        
        \cmidrule(lr){1-7}
        
        \multicolumn{7}{l}{Context-aware Ranking Models}  \\
        
        \cmidrule(lr){1-7}
        
        CARS & 0.7385$^\dag$ & 0.7386$^\dag$ & 0.7512$^\dag$ & 0.8837$^\ddag$ & 0.8556$^\ddag$ & 0.9268$^\ddag$ \\
        
        HBA & 0.7612$^\ddag$ & 0.7518$^\dag$ & 0.7639$^\dag$ & 0.8896$^\ddag$ & 0.8615 & 0.9316$^\ddag$ \\ 
        
        RICR & 0.7670$^\ddag$ & 0.7636$^\ddag$ & 0.7740$^\ddag$ & 0.8934$^\ddag$ & 0.8147$^\dag$ & 0.8937$^\dag$ \\
        
        HQCN & 0.7739$^\ddag$ & {0.7682} & {0.7783} & {0.8976} & 0.8659 & 0.9328$^\ddag$ \\ 
        
        BART & {0.7380$^\dag$} & {0.7464$^\dag$} & {0.7574$^\dag$} & {0.8853$^\ddag$} & {0.8585$^\ddag$} & {0.9294$^\ddag$} \\ 

        BERT & {0.7488$^\dag$} & {0.7541$^\ddag$} & {0.7651$^\dag$} & {0.8890$^\ddag$} & {0.8653} & {0.9316$^\ddag$} \\ 
        
        COCA & {0.7769} & 0.7576$^\ddag$ & 0.7703$^\ddag$ & 0.8932$^\ddag$ & 0.8623 & {0.9382} \\
        
        {ASE} & \textbf{0.7884} & \textbf{0.7727} & \textbf{0.7839} & \textbf{0.8996} & \textbf{0.8701} & \textbf{0.9482} \\
    \bottomrule
    \end{tabular}
    \vspace{-5px}
    \label{tab:result}
\end{table}

The overall performances of all models are presented in Table~\ref{tab:result}. 
The results show that context-aware ranking models generally perform better than ad-hoc ones, which indicates the effectiveness of modeling session context.
Besides, we can further obtain the following observations:

\textbf{(1) ASE outperforms all baselines in terms of all metrics on both datasets.} 
Specifically, ASE outperforms COCA, a strong baseline that utilizes pre-training and data augmentation strategies.
It demonstrates the effectiveness of utilizing our generative tasks to model the current search intent.
In future work, we will try to incorporate pre-training techniques (\eg, contrastive learning) and data augmentation strategies (\eg, curriculum learning) into ASE to further improve its performance.
Besides, we will notice that the improvements of ASE on AOL are more significant than those on Tiangong-ST.
The potential reasons are as follows:
(i) The base performances on Tiangong-ST are already very high because there are more than 77.4\% candidate documents with relevance scores that are larger than 1 in the test set.
Specifically, even the BM25 algorithm can achieve 0.8541 in terms of NDCG@10 on this dataset.
Therefore, it is more difficult for ASE to improve its performance on Tiangong than AOL.
This phenomenon has also been noticed by Zhu et al.~\cite{coca}.
(ii) As suggested by the authors of this dataset~\cite{tiangong}, we use the queries that have human-annotated relevance labels when testing, and all of them are the last ones in their sessions.
However, as we find in Section~\ref{subsec:session_len}, ASE can give more considerable improvements on the queries with fewer histories.
This is because ASE utilizes future sequences and a supplemental query to train the encoder to predict the actual search intent.

\textbf{(2) The vanilla version of the BART model underperforms that of BERT.} 
For example, BERT and BART achieve about 0.3990 and 0.3908 in terms of NDCG@1 on AOL dataset, respectively.
This demonstrates that the original encoder of BART performs worse than BERT's encoder, which makes the comparisons of ASE and BERT-based baselines (HBA, COCA) fair (as illustrated in Section~\ref{subsec:overview}).
We believe the reason is that BERT-base has 12 layers, whereas the encoder of BART-base only has 6 layers.
However, ASE can still outperform the BERT-based baselines based on a worse backbone than BERT (for the encoder), which further demonstrates its effectiveness.

\begin{table}[t!]
    \centering
    \small
    \caption{Performances of ablated models on AOL dataset.}
    \begin{tabular}{l|c|c|c|c}
    \toprule
         Metric & w/o. PFQ & {w/o. PCD} & w/o. PSQ & ASE \\
        \midrule
        
        NDCG@1 & 0.4100 \ \ -1.06\% & {0.4036 \ \ -2.61\%} & 0.4102 \ \ -1.01\%  & \textbf{0.4144} \\
        
        NDCG@3 & 0.5580 \ \ -1.80\% & {0.5570 \ \ -1.97\%}  & 0.5636 \ \ -0.81\%  & \textbf{0.5682} \\
        
        NDCG@5 & 0.5933 \ \ -1.23\% & {0.5895 \ \ -1.86\%} & 0.5957 \ \ -0.83\%  & \textbf{0.6007} \\
        
        NDCG@10 & 0.6205 \ \ -1.24\% & {0.6180 \ \ -1.64\%}  & 0.6246 \ \ -0.59\%  &  \textbf{0.6283} \\
        
        MAP & 0.5579 \ \ -1.26\% & {0.5546 \ \ -1.84\%}  & 0.5608 \ \ -0.74\%  & \textbf{0.5650} \\
        
        MRR & 0.5691 \ \ -1.06\% & {0.5650 \ \ -1.77\%} & 0.5707 \ \ 
        -0.78\%  & \textbf{0.5752} \\
        
    \bottomrule
    \end{tabular}
    \vspace{-5px}
    \label{tab:ablation}
\end{table} 

\subsection{Ablation Studies} \label{subsec:ablation}

To demonstrate the effectiveness of the generative tasks for helping the ranking task, we design several variants of ASE. 
Specifically, we conduct ablation experiments on AOL dataset as follows:

$\bullet$ \textbf{ASE w/o. PFQ.} We remove the task of Predicting Future Queries (PFQ).

$\bullet$ { \textbf{ASE w/o. PCD.} } We discard the task of Predicting future Clicked Documents (PCD).

$\bullet$ { \textbf{ASE w/o. PSQ.} } We abandon the task of Predicting a Supplemental Query (PSQ).

The performances are presented in Table~\ref{tab:ablation}. 
All the ablated models perform worse than the full ASE, which demonstrates the effectiveness of utilizing our generative tasks to model current search intent.
Specifically, we can draw these conclusions:

\textbf{(1) Predicting future queries is effective for inferring the actual search intent.}
In Section~\ref{subsec:generate}, we propose to treat future queries as a generation target because they have higher quality than the current one.
After removing this task, ASE's performance drops.
Specifically, it drops about 1.80\% in terms of NDCG@3.
This demonstrates the effectiveness of this task.

\textbf{(2) Predicting future clicked documents can help the ranking task.}
In Section~\ref{subsec:generate}, we propose to treat future clicked documents as another generation target because they are more accurate representations of search intent.
After discarding this task, ASE's performance decreases.
Specifically, it decreases about 2.61\% in terms of NDCG@1.
This indicates that utilizing the information of future clicked documents can help the ranking task.

\textbf{(3) Predicting a supplemental query can make our model more robust.}
In Section~\ref{subsec:generate}, we attempt to mine a supplemental query from other sessions to supplement our understanding of the current query.
After abandoning this task, ASE's performance declines.
For example, it declines by about 1.01\% in terms of NDCG@1.
This shows that this task can make our model more robust.

\begin{table}[t!]
    \centering
    \small
    \caption{Performances of different generative targets on AOL dataset.
    Suppose the current session sequence is $S = \{q_1, d_1, q_2, d_2, \cdots , q_n\}$. 
    } 
    \begin{tabular}{l|c|c|c}
    \toprule
    
        $GT$ & NDCG@1 & NDCG@10 & MAP  \\ 
        
        \midrule
        
        - (BART)  &  {0.3882} &  {0.6124} &  {0.5450} \\
        
        \midrule
        
        $q_{n-1}$ & {0.3849} \ \ -0.85\% & {0.6103} \ \ -0.34\% & {0.5427}\ \ -0.42\%  \\
        
        \midrule
        
        $q_{n}$ & {0.3928} \ \ +1.84\% & {0.6077} \ \ -0.77\% & {0.5442}\ \ -0.15\% \\
        
        \midrule
        
        $q_{n+1}$ & {0.4004} \ \ +3.14\% & {0.6150} \ \ +0.42\% & {0.5516}\ \ +1.21\% \\
        
        \midrule
        
        $d_{n-1}$ & {0.3922} \ \ +1.03\% & {0.6104} \ \ -0.33\% & {0.5464}\ \ +0.26\%  \\
        
        \midrule
        
        $d_{n}$ & {0.4022} \ \ +3.61\% & {0.6212} \ \ +1.44\% & {0.5548}\ \ +1.80\% \\
        
        \midrule
        
        $d_{n+1}$ & {0.4044} \ \ +4.17\% & {0.6206} \ \ +1.34\% & {0.5565}\ \ +2.11\% \\
        
        \midrule
        
        $q_n'$ & {0.3990} \ \ +2.78\% & {0.6151} \ \ +0.44\% & {0.5509}\ \ +1.08\% \\

    \bottomrule
    \end{tabular}
    \label{tab:generations}
\end{table}

\subsection{Performances of Various Generative Targets} \label{subsec:explore_generation}

Given the current sequence $S = \{q_1, d_1, q_2, d_2, \cdots , q_n\}$, we explore extensive possibilities of generation targets, \eg, preceding queries, the current query, subsequent queries, historical clicked documents, future clicked documents, and a supplemental query. 
We will treat each of them \textbf{as the only generative task} here and jointly learn to generate it with the ranking task. 
The performances of these generative targets on AOL dataset are presented in Table~\ref{tab:generations}.
Note that we only use one sequence per $GT$ to get a straightforward view of its effectiveness. 
In the following section, we will study the length of $GT$s (\ie, the prediction window size $w$).
From these results, we can draw the following conclusions:

\textbf{(1) Predicting the future sequences are more effective than simply recovering the historical ones.}
As explained in Section~\ref{subsec:generate}, the future queries and documents can represent the search intent.
As presented in Table~\ref{tab:generations}, we can notice that predicting the future behaviors achieves more considerable improvement over the ranking task than recovering the historical behaviors.
Specifically, predicting the following query $q_{n+1}$ increases the performance of the ranking task by about 1.21\% in terms of MAP on AOL, whereas recovering the previous query $q_{n-1}$ makes the performance drop by about 0.42\% in terms of MAP.

\textbf{(2) It is generally more helpful to predict the information of clicked documents than predict queries.}
As stated in Section~\ref{subsec:generate}, the clicked documents can often represent the search intent more accurately.
As shown in Table~\ref{tab:generations}, treating the clicked documents as generative targets perform better than queries.
Specifically, predicting the current clicked document $d_n$, and the following query $q_{n+1}$ increase the performance of BART by about 3.61\% and 3.14\% in terms of NDCG@1 on AOL, respectively.
We compare the results of $d_n$ and $q_{n+1}$ because they are considered the first behaviors in the subsequent sequences of $S$.

\textbf{(3) Predicting a supplemental query can help the ranking task.}
As illustrated in Section~\ref{subsec:generate}, a supplemental query can supplement the understanding of the search intent.
As shown in the last line of Table~\ref{tab:generations}, predicting a supplemental query $q_n'$ increases the performance by about 1.08\% in terms of MAP on AOL.
This indicates that our third generative task can help the ranking task.

\begin{figure}
    \centering
    \includegraphics[width=0.8\linewidth]{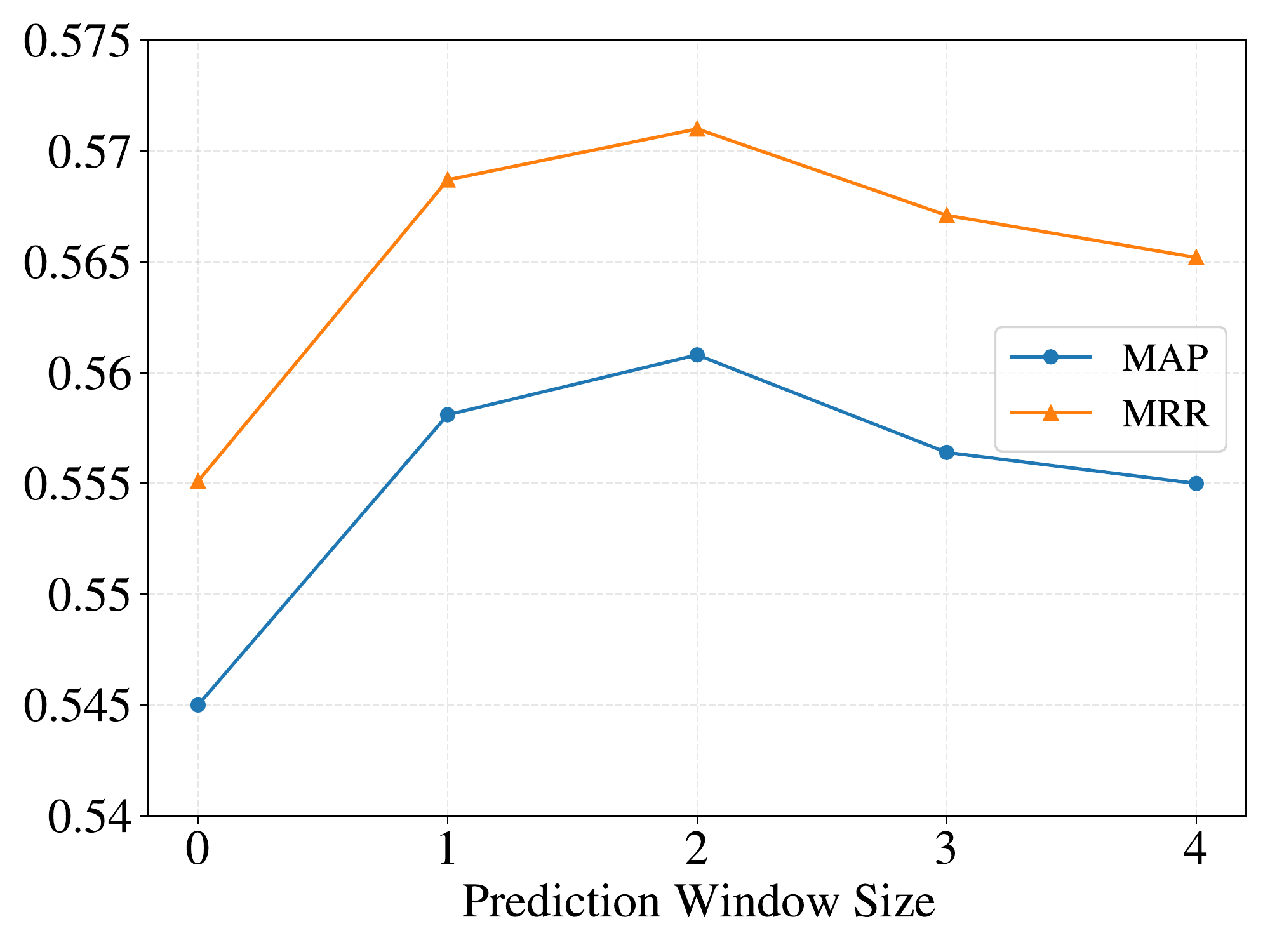}
    \caption{Performances of the variant of ASE (BART with $GT_1$ and $GT_2$) with different values of $w$ on AOL dataset.}
    \label{fig:window}
\end{figure}

\subsection{Effect of Prediction Window Size} \label{subsec:pred_window}

In Section~\ref{subsec:generate}, we try to utilize the information of future sequences to help encode the current sequence. 
Specifically, we treat the future queries and clicked documents as generation targets in Task 1 and Task 2.
We use a prediction window size $w$ to control the number of subsequent behaviors to generate.
To determine the value of $w$, we finetune a variant of ASE (BART with $GT_1$ and $GT_2$) under different settings of $w$ on the validation sets. 
We do not include Task 3 ($GT_3$) in this variant because we want to directly estimate $w$'s influence on the ranking task without the effect of $GT_3$.
We find that our model performs best when $w$ is 2 on the validation sets. 
In Figure~\ref{fig:window}, we show the performances of this variant under different $w$ on the test set of AOL.
Note that we show the results of the test set only for consistency with previous experiments' results.
$w$ is tuned based on performances on the validation sets.

From Figure~\ref{fig:window}, we can find the performance increases from 0 to 2 and slowly decreases from 2 to 4.
We believe there is a trade-off.
If $w$ is too small (0,1), the encoder can not actually encode the search intent into the high-level representations.
And if $w$ is too large (3,4), the generation target may become too hard to generate. Besides, the average session lengths are about 2.5 for both datasets (Table~\ref{tab:dataset}), so there will be many empty sequences in the $GT$s if $w$ is too large.

\subsection{Application to Other Transformer-based Encoder-Decoder Models} \label{subsec:other_plms}

\begin{figure*}[t!]
    \centering
    \begin{subfigure}[b]{0.495\linewidth}
        \centering
        \includegraphics[width=\linewidth]{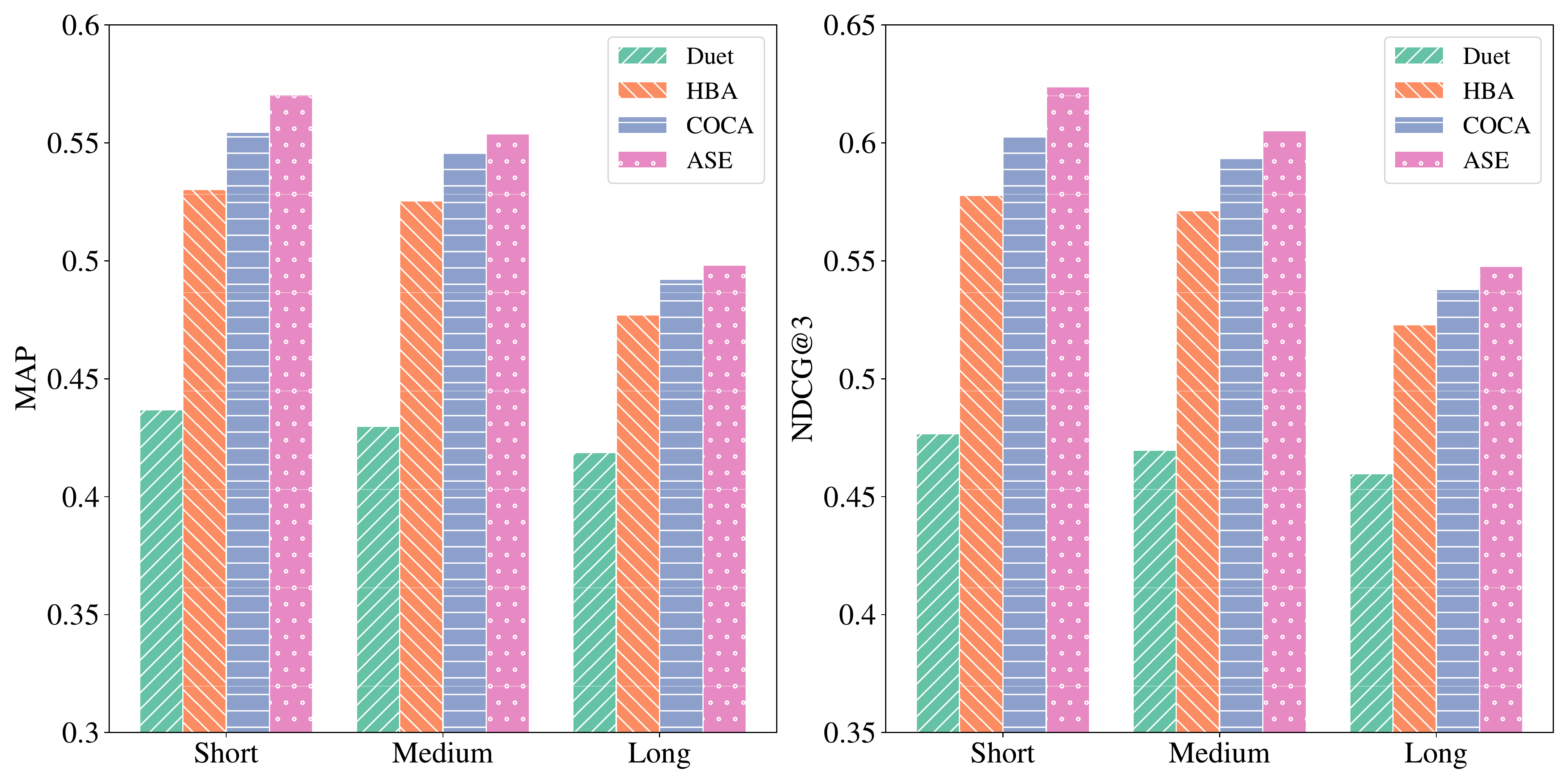}
    \end{subfigure}
    \begin{subfigure}[b]{0.495\linewidth}
        \centering
        \includegraphics[width=\linewidth]{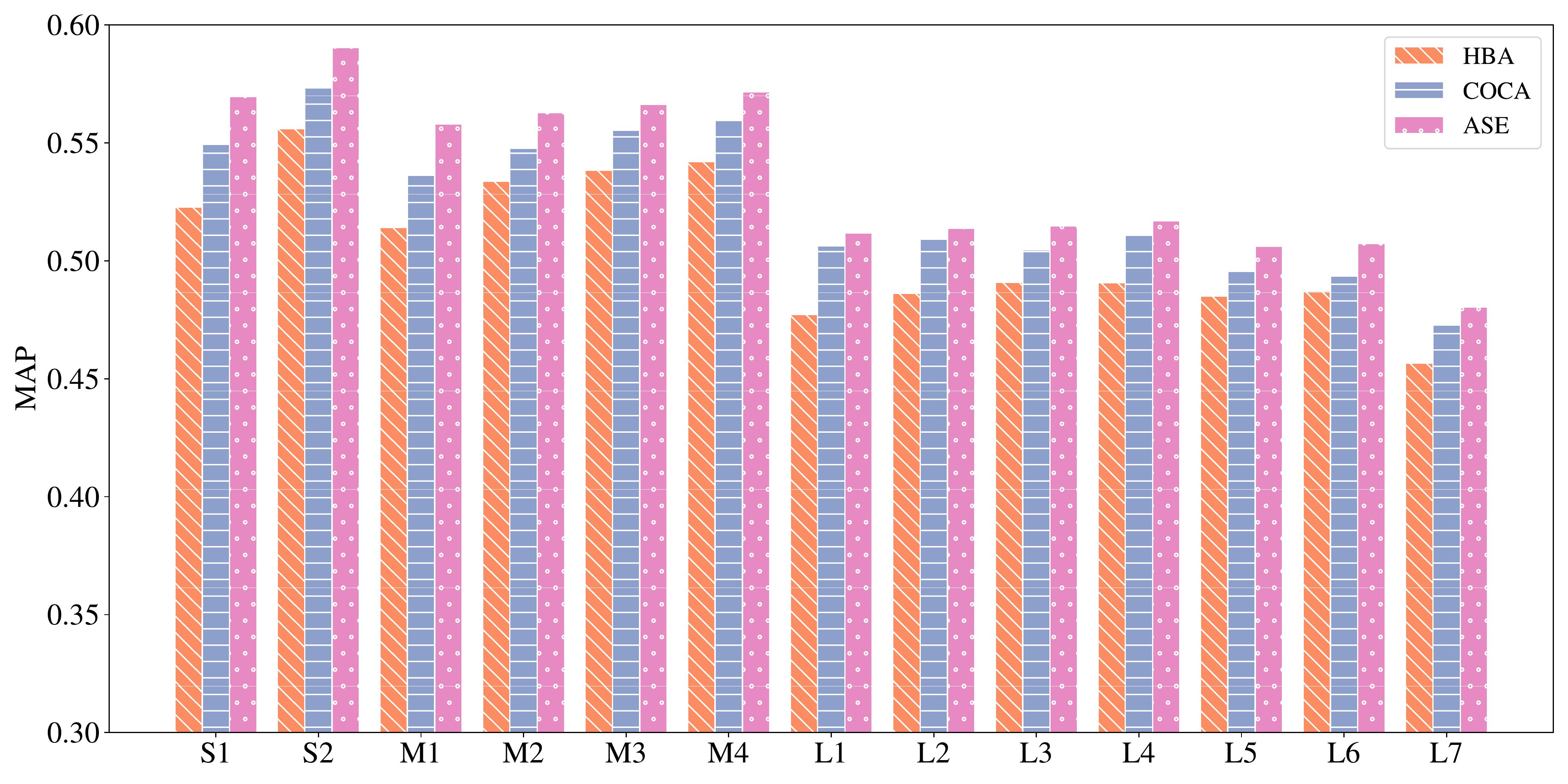}
    \end{subfigure}
    \caption{ The left part presents the performance comparison of HBA, COCA, and ASE on sessions with different lengths on AOL dataset. 
    The right part shows the performance comparison of Duet, HBA, COCA, and ASE at different query positions in short (S1-S2), medium (M1-M4), and long sessions (L1-L7).
    The number after ``S'', ``M'', or ``L'' indicates the query index in a task.}
    \label{fig:session_len}
\end{figure*}

\begin{table}[t!]
    \centering
    \small
    \caption{ The performances of the base models and the models with our generative tasks (+$GT$s) on AOL dataset.
    ``$\dag$'' indicates the result is significantly worse than the model with $GT$s in t-test with $p$-value $<$ 0.01. } 
    \begin{tabular}{lcccc}
    \toprule

        \textbf{Model} & MAP & MRR & NDCG@3 & NDCG@10 \\ 
        
        \midrule
        
        T5-small & 0.5142$^\dag$ & 0.5257$^\dag$ & 0.5102$^\dag$ & 0.5803$^\dag$ \\
        
        T5-small + $GT$s & 0.5246 & 0.5363 & 0.5232 & 0.5911 \\
        
        Improv. & +2.02\% & +2.02\% & +2.55\% & +1.86\% \\
        
        \midrule
        
        BlenderBot-small & 0.5465$^\dag$ & 0.5570$^\dag$ & 0.5470$^\dag$ & 0.6108$^\dag$\\

        BlenderBot-small + $GT$s & 0.5580 & 0.5685 & 0.5601 & 0.6220\\
        
        Improv. & +2.10\% & +2.06\% & +2.39\% & +1.83\% \\
 
    \bottomrule
    \end{tabular}
    \label{tab:other_plms}
\end{table}

As illustrated in Section~\ref{subsec:overview}, we choose BART as our model's backbone mainly for fair comparisons with BERT-based baselines (HBA, COCA).
However, our approach can be easily applied to other Transformer-based encoder-decoder structured models.
In this section, we choose two seq2seq models (T5~\cite{t5} and BlenderBot~\cite{blender}) as the base models.
For T5, we use the small version provided by Huggingface.\footnote{\url{https://huggingface.co/t5-small}}
For BlenderBot, we use the small version provided by Facebook on Huggingface.\footnote{\url{https://huggingface.co/facebook/blenderbot\_small-90M}}
They are fine-tuned with the ranking task that is the same as BERT's and BART's strategy introduced in Section~\ref{subsec:baselines}.
We also train them with our designed generative tasks, and the corresponding results are reported as ``X+$GT$s''.

As presented in Table~\ref{tab:other_plms}, the models with our designed generative tasks outperform their base models significantly on AOL dataset, respectively.
Specifically, T5-small model with $GT$s improves T5-small by more than 2.02\% in terms of MAP.
This indicates that utilizing our generative tasks to enhance session context modeling is effective under different backbones, and our approach can be easily applied to other Transformer-based encoder-decoder structured models than BART.

\subsection{Performance on Different Query Positions and Sessions with Different Lengths} \label{subsec:session_len}

Following previous works~\cite{cars, hqcn, ricr, coca}, in order to study ASE's performance on sessions with different lengths, we split the test dataset of AOL as follows:

$\bullet$ Short sessions (with 2 queries) - 66.5\% of the test set.

$\bullet$ Medium sessions (with 3-4 queries) - 27.24\% of the test set.

$\bullet$ Long sessions (with 5+ queries) - 6.26\% of the test set.

We compare ASE with Duet, HBA, and COCA on these different sessions. 
The results are presented in the left part of Figure~\ref{fig:session_len}.
We can find that: 
(1) The ad-hoc ranking model Duet performs worse than context-aware models, which indicates the importance of modeling session context.
(2) ASE outperforms other models on all lengths of sessions, which demonstrates the effectiveness of utilizing our generative tasks to model session context.
(3) ASE performs worse on long sessions than on short sessions.
As explained in~\cite{cars, coca}, long sessions are intrinsically more difficult.
The similar declining trends of other models also demonstrate this idea.

To study ASE's performance of modeling task progression, we also compare it with HBA and COCA on different query positions.
The results are shown in the right part of Figure~\ref{fig:session_len}.
We can find that the performances most increase as the session progresses because there is more session context to model.
However, compared to COCA, ASE has a relatively slower speed for improvement (or performs better on queries that lack context). 
This is because ASE utilizes our generative tasks during training, which can help its encoder predict the search intent even with few historical behaviors.
Besides, it is interesting that all models' performances decrease from L4 to L7.
We believe these long sessions often represent complex or exploratory search tasks, which are hard to complete. 

\section{Conclusions and future work} \label{sec:conclusion}

In this work, we attempt to utilize generative tasks to model session context.
An encoder-decoder structure and three generative tasks are used to enhance the ability of the encoder.
With these generative tasks, we aim to train our model to predict future queries, future clicked documents, and a supplemental query.
We believe that if our model could predict these sequences, then the actual search intent has been successfully encoded into the high-level representations of the current session sequence.
Rich experiments on two public search logs demonstrate the effectiveness and broad applicability of our approach.

Nevertheless, our work still has some limitations that we plan to address in future work: 
(1) Though ASE outperforms COCA without pre-training and data augmentation strategies, ASE still has potential variants that incorporate pre-training techniques.
In future work, we will try to incorporate pre-training techniques (\eg, contrastive learning) and data augmentation strategies (\eg, curriculum learning) into ASE to further improve its performance.
(2) The method of mining a supplemental query from the database is relatively naive.
We plan to use more sophisticated algorithms or models (\eg, Sentence Transformer) to find a query with higher quality.
(3) In order to further denoise the current session, the historical behaviors may be treated with distinction.
For example, we could first extract or generate keywords from the behaviors before putting them into the encoder.

\begin{acks}

Zhicheng Dou is the corresponding author. Ji-Rong Wen is also with Key Laboratory of Data Engineering and Knowledge Engineering, MOE.
This work was supported by the National Natural Science Foundation of China No. 61872370 and No. 61832017, Beijing Outstanding Young Scientist Program NO. BJJWZYJH012019100020098, and Intelligent Social Governance Platform, Major Innovation \& Planning Interdisciplinary Platform for the ``Double-First Class'' Initiative, Renmin University of China.

\end{acks}

\balance

\end{document}